# Hidden Magnetism and Quantum Criticality in the Heavy Fermion Superconductor CeRhIn$_5$

Tuson Park*, F. Ronning*, H. Q. Yuan†, M. B. Salamon†, R. Movshovich*, J. L. Sarrao*, & J. D. Thompson*

*Los Alamos National Laboratory, Los Alamos, NM 87545, USA

†Department of Physics, University of Illinois at Urbana-Champaign, Urbana, IL 61801, USA

**With understood exceptions, conventional superconductivity does not coexist with long-range magnetic order[1]. In contrast, unconventional superconductivity develops near a boundary separating magnetically ordered and magnetically disordered phases[2,3]. A maximum in the superconducting transition temperature $T_c$ develops where this boundary extrapolates to $T$=0 K, suggesting that fluctuations associated with this magnetic quantum-critical point are essential for unconventional superconductivity[4,5]. Invariably though, unconventional superconductivity hides the magnetic boundary when $T < T_c$, preventing proof of a magnetic quantum-critical point[5]. Here we report specific heat measurements of the pressure-tuned unconventional superconductor CeRhIn$_5$ in which we find a line of quantum-phase transitions induced inside the superconducting state by an applied magnetic field. This quantum-critical line separates a phase of coexisting antiferromagnetism and superconductivity from a purely unconventional superconducting phase and terminates at a quantum tetracritical point where the magnetic field completely suppresses superconductivity. The $T\rightarrow 0$ K magnetic field-pressure phase diagram of CeRhIn$_5$ is well described with a theoretical model[6,7] developed to explain field-induced magnetism in the high-$T_c$ cuprates but**

**in which a clear delineation of quantum-phase boundaries has not been possible. These experiments establish a common relationship among hidden magnetism, quantum criticality and unconventional superconductivity in cuprate and heavy-electron systems, such as CeRhIn$_5$.**

CeRhIn$_5$ belongs to a family of cerium, uranium and plutonium-based compounds in which electronic interactions enhance the effective mass of charge carriers up to 1,000 times the mass of a free electron. These heavy-electron materials in turn belong to a larger family of strongly correlated electron systems, which include the high-T$_c$ cuprate superconductors and some organic compounds. A generic temperature-control parameter ($T$-$\delta$) phase diagram common to strongly correlated unconventional superconductors is shown in Fig. 1 (Ref. 5, 8-10). CeRhIn$_5$ is prototypical of this phase diagram, in this case with pressure as the tuning parameter[11,12]. At atmospheric pressure, CeRhIn$_5$ orders antiferromagnetically at 3.8 K, and with applied pressure the antiferromagnetic state vanishes at $P_{c1}$ = 1.77 GPa when the Neel temperature $T_N$ equals the superconducting transition temperature $T_c$ = 1.9 K. Over a range of pressures below $P_{c1}$, extensive measurements show[12-15] that magnetic order coexists with superconductivity, but only when $T_N > T_c$. Above $P_{c1}$, where $T_c > T_N$ (extrapolated), these same measurements find only unconventional superconductivity. A smooth extrapolation of $T_N$ ($P$) to $T = 0$ K suggests that Néel order, if it existed, would terminate at a quantum critical point (QCP) $P_{c2}$ near 2.3 GPa, where the effective mass of charge carriers diverge in the normal state.[16]

Figure 2 summarizes field-dependent specific heat measurements on a single crystal of CeRhIn$_5$ subjected to pressures just below and above $P_{c1}$. For $P$ = 1.68 GPa < $P_{c1}$ (panel a) these data confirm earlier conclusions[12-18] that AFM order coexists with SC. Above $P_{c1}$ (panel b), only a specific heat discontinuity due to SC is observed for fields up to 44 kOe: there is no evidence for a magnetic phase transition at these low fields, consistent with other measurements at zero applied field[12-15,19]. At 55 kOe, a specific-heat anomaly near 0.7 K emerges below the SC transition ($T_c$ = 1.7 K) and grows in intensity with increasing field. Finally, in the bottom panel ($P$ = 2.3 GPa), there is no signature for a magnetic transition up to 88 kOe and down to 300 mK. As the superconducting transition is suppressed to zero, $C/T$ diverges weakly with decreasing temperature.

Figure 3a shows the evolution of the field-induced magnetic anomaly in $C/T$ for $P$ = 2.1 GPa. Similar results were obtained at $P$ = 1.8 and 1.9 GPa, even closer to $P_{c1}$. The area under these curves is a measure of magnetic entropy, plotted in the inset, and reflects approximately the magnitude of the field-induced magnetism. The near linear proportionality of the entropy to the applied field suggests that magnetism is associated with quantized vortices of magnetic flux that penetrate the superconductor and whose areal density is proportional to $H$. The $H$-induced transition temperature increases with $H$, consistent with intrinsic magnetism, not with superconducting nor extrinsic phases. An explanation for these observations is discussed later.

A temperature-pressure phase diagram constructed from specific heat measurements is plotted in Fig. 3b for representative fixed fields. For zero magnetic field, evidence for a magnetic transition temperature abruptly disappears at $P_{c1}$, suggesting a first-order like transition. In an applied field of 33 kOe, however, the line of second order magnetic transition temperatures $T_{N1}$ smoothly evolves through $P_{c1}$ deep into the SC dome. With increasing field, the relative position of the critical points where $T_{N1}$ becomes zero changes with respect to the centre of the SC dome. The influence of superconductivity on the development of magnetic order is reflected in a slope change of the magnetic transition line as it crosses into the superconducting domain.

These results are shown in Fig. 4 as a temperature-pressure-field phase diagram. The vertical $H$-$P$ plane at a fixed temperature of 0.65 K changes very little with decreasing temperature to 350 mK, and we take it as representative of a $T = 0$ K plane (see Supplementary Fig. 1). With increasing pressure, the magnetic field required to induce magnetism increases and finally meets with the upper critical field line ($H_{c2}$) at $P_{c2}$ (= 2.25 GPa), a tetracritical point that branches out a transition line between magnetically ordered and disordered phases for $H > H_{c2}$. This observation of the $H$-induced magnetism only for $P_{c1} < P < P_{c2}$ provides now an explanation for the dHvA observation[16] of a diverging effective mass as due to a field- and pressure-tuned QCP at $P_{c2}$. At this pressure, the Fermi surface volume expands to accommodate additional delocalised charge carriers. The larger Fermi volume of CeRhIn$_5$ beyond $P_{c2}$ corresponds closely to that of the isostructural superconductor CeCoIn$_5$ whose 4f electron from cerium contributes to the Fermi volume[16]. A localized to delocalised transition in the 4f-electron configuration is expected in a model of criticality in which extended and localized fluctuations coexist at a quantum critical point[20]. This model, however, does not include the role of superconductivity.

Neutron-diffraction experiments also have revealed field-induced magnetic order in the superconducting state of the high-$T_c$ compound La$_{1.9}$Sr$_{.1}$CuO$_4$ (Ref. 21). Motivated by these observations, Demler *et al.* proposed a model that assumes the superconductor is near a quantum phase transition to a state with microscopic coexistence of superconducting and magnetic orders[6,7]. When the magnetic field penetrates an unconventional superconductor in which the SC energy gap has nodes on the Fermi surface, field-induced quantized vortices have AFM as their ground state, which suppresses superconductivity around the vortices. The suppression of the SC order enhances the competing AFM order even outside of the normal vortex cores, thus delocalising magnetic correlations and creating microscopic coexistence of AFM and SC. Repulsive coupling between AFM and SC orders, which can be tuned either by chemical substitution or pressure, tips the balance between the two competing grounds states, leading to a quantum phase transition among pure AFM phase, AFM+SC coexisting phase, and pure SC phase. This model accounts for the evolution of magnetic order in $La_{1.9}Sr_{.1}CuO_4$ and its strengthening with increasing field[21].

This model[6,7] further predicts a line of quantum phase transitions between the AFM+SC and SC phases as a function of a control parameter $\delta$. Taking pressure as the control parameter, this model predicts: $H/H_{c2}^0 \approx 1-\gamma[1-\alpha(P-P_{c1})]$ for $H/H_{c2}^0 > 0.1$, where $H_{c2}^0$ is the upper critical field at a tetracritical point ($P_{c2}$ in Fig. 4), $\gamma$ is a numerical constant, and $\alpha$ is a proportionality between pressure and a repulsive coupling constant ($\delta = \alpha P$). A least-squares fit of this relationship to the open squares in Fig. 4 gives $\alpha = 1.99$, $\gamma = 1.11$, and $P_{c1} = 1.75$ GPa (dashed line in the $H$-$P$ plane of Fig. 4). The numerical constant $\gamma$ is in good agreement with that ($=1.2$) obtained from a numerical solution of this model[7]. We also obtain $P_{c2}$ ($H = H_{c2}^0$) = 2.25 GPa, which is very close to the pressure where the effective mass of charge carriers diverges[16]. The $H$-linear proportionality of the magnetic entropy (inset to Fig. 3a) and the microscopic coexistence of AFM and SC in $CeRhIn_5$ are consistent with this model[6,7], which considers AFM order as the competing ground state of SC . A possible explanation for the above phenomenological description of $CeRhIn_5$ is that the presence of SC strongly inhibits a mechanism by which spins communicate, such as the Rudermann-Kittel-Kasuya-Yoshida (RRKY) interaction, which then may explain why magnetism is hidden in zero field by superconductivity when $T_c > T_N$.

Similarities between the high-$T_c$ cuprates[21-25] and $CeRhIn_5$ suggest that phenomena in them may be ubiquitous features of magnetically mediated superconductivity. The model of criticality that accounts for our data is not specific to the microscopic origin of unconventional superconductivity or of quantum criticality. Whereas, AFM is due to localized 4f electrons and quantum criticality is associated with a localized to delocalised transition in the 4f configuration of $CeRhIn_5$, this is not an appropriate description of cuprate physics nor possibly of all heavy-electron compounds.

Consequently, within this model, the mechanism of unconventional superconductivity may differ in detail among systems, even though magnetism is a common denominator.

**Supplementary Information** is linked to the online version of the paper at www.nature.com/nature. Experimental methods and a *H-T* phase diagram for $CeRhIn_5$ at various pressures are available in SI.

**Acknowledgements** The authors thank Y. K. Bang, A. V. Balatsky and N. J. Curro for discussions. Work at Los Alamos National Laboratory was performed under the auspices of the US Department of Energy, Office of Science. HQY acknowledges the ICAM postdoctoral fellowship.

**Author Information** Reprints and permissions information is available at npg.nature.com/reprintsandpermissions. The authors declare no competing financial interests. Correspondence and requests for materials should be addressed to T. P. (tuson@lanl.gov).

**Figure Legends**

Figure 1. Schematic temperature-control parameter ($T$-$\delta$) phase diagram common to classes of unconventional superconductors. The hatched area represents an antiferromagnetically (AFM) ordered state and the colored area, a superconducting (SC) phase. The hatched area in the colored background denotes coexisting AFM and SC. In the normal state above the SC dome, physical properties are not typical of a metal and reflect non-Fermi liquid (NFL) behaviours. As the control parameter, such as chemical substitution or pressure is varied, long-range magnetic order gives way to a superconducting state. Above the superconducting dome, normal state properties are dominated by long-ranged, long-lived fluctuations that are expected if the magnetic phase boundary extended smoothly to absolute zero temperature, i.e., to a magnetic quantum critical point ($\delta_2$). Experimentally, however, magnetic order abruptly disappears at a finite temperature where the superconductivity and magnetic phase boundaries meet, suggesting a first order or weakly first order boundary at $\delta_1$ and providing no obvious connection between magnetism and the putative $\delta_2$. In such a case, it is difficult to reconcile the existence of an extended range of unconventional superconductivity beyond $\delta_1$ and of an unusual normal state above $T_c$.

Figure 2. Specific heat divided by temperature, $C/T$, as a function of temperature for CeRhIn$_5$ at fixed magnetic fields. The magnetic field is applied perpendicular to the *c*-axis of this tetragonal compound. Specific heat is determined by an ac calorimetric method (see Supplementary Methods). a) At 1.68 GPa, specific heat anomalies due to antiferromagnetic transitions $T_{N1}$ and $T_{N2}$ and a superconducting transition $T_c$ are observed. $T_{N1}$ signals the onset of incommensurate antiferromagnetism with propagation wave vector (½, ½, 0.297) (Ref. 17) and $T_{N2}$ = 1.85 K is due to a spin reorientation transition[18]. As at atmospheric pressure, $T_{N1}$ and $T_{N2}$ are almost independent of the applied field. A specific heat discontinuity due to superconductivity follows at 1.55 K. b) At 2.1 GPa, only a superconducting anomaly appears for $H \leq 44$ kOe (diamonds). At 55 kOe (side triangles), however, magnetism appears for $T < T_c(H)$. With further increasing field, the magnetic anomaly is enhanced and persists for $T > T_c(H)$, as shown in Supplementary Fig. 1. c) At 2.3 GPa, only SC appears and magnetism is absent for $H \leq 88$ kOe and $T > 300$ mK.

Figure 3. a) Field-dependence of the magnetic specific heat of CeRhIn$_5$ at 2.1 GPa. To estimate the magnetic contribution to the specific heat $\Delta C/T$ due to magnetic order, we assume a smoothly varying background ($\sim T^2$) over a temperature range of interest and subtract that background contribution from the total measured specific heat. Inset: Entropy involved in the magnetic ordering transition, i.e., area under the peak in $\Delta C/T$, as a function of magnetic field. The dashed line is a guide to eyes. The relative entropy associated with magnetic order is much less for $P$ = 2.1 GPa than at $P$ = 0 for any $H$. b) Temperature-pressure diagram at $H$ = 0 (squares), 33 (circles), and 88 kOe (triangles). The magnetic transition is depicted as solid symbols and the superconducting transition by open symbols. Lines are spline fits to the data points.

Figure 4. *H-T-P* phase diagram of the heavy fermion superconductor $CeRhIn_5$. The temperature-pressure plane is at $H$ = 0 kOe and field-pressure plane is for $T$ = 650 mK. In the *T-P* plane, the magnetically ordered (MO) state of Ce 4f moments is preferred at low pressure. With increasing pressure, a superconducting (SC) phase appears and coexists with the MO phase when $P$ < 1.77 GPa. For 1.77 < $P$ < 2.3 GPa, only a SC phase is found in zero field, but applied magnetic field induces a MO phase in the SC state. The blue line is a proposed pseudogap line[13]. In the *H-P* plane, upper critical fields $H_{c2}$, where superconductivity is totally depressed due to the overlap of magnetic vortex cores, are represented by green circles. Quantum phase transitions between the pure SC phase and the coexistence phase of *H*-induced magnetism and SC are shown as open squares. The hatched grey line delineates a boundary between the MO phase and a magnetically disordered (MD) phase in the normal state. This boundary is defined by field-dependent specific heat and dHvA measurements, which were made at milliKelvin temperatures and fields 88 < $H$ < 169 kOe (Ref. 16). $P_{c1}$ is a quantum phase transition point between SC+MO and SC phases at zero magnetic field. $P_{c2}$ is a tetracritical point where the $H_{c2}$ line and the MO to MD lines cross. Experimental data constrain $P_{c2}$ to within $\pm 0.05P$. The dashed line between $P_{c1}$ and $P_{c2}$ is a fit to the data as described in the text.

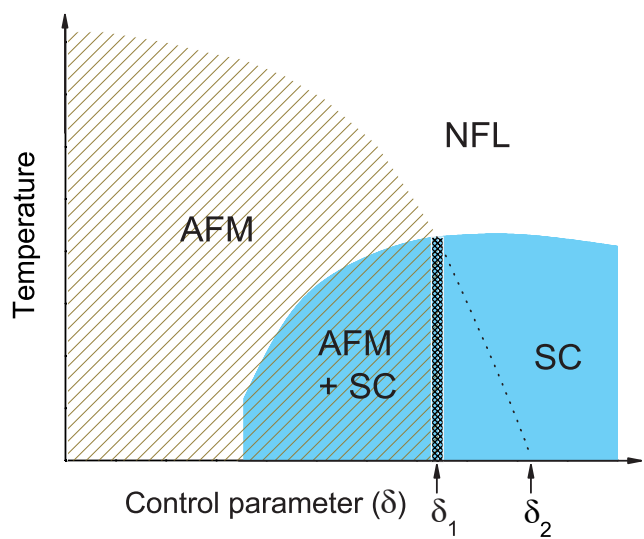

Figure 1

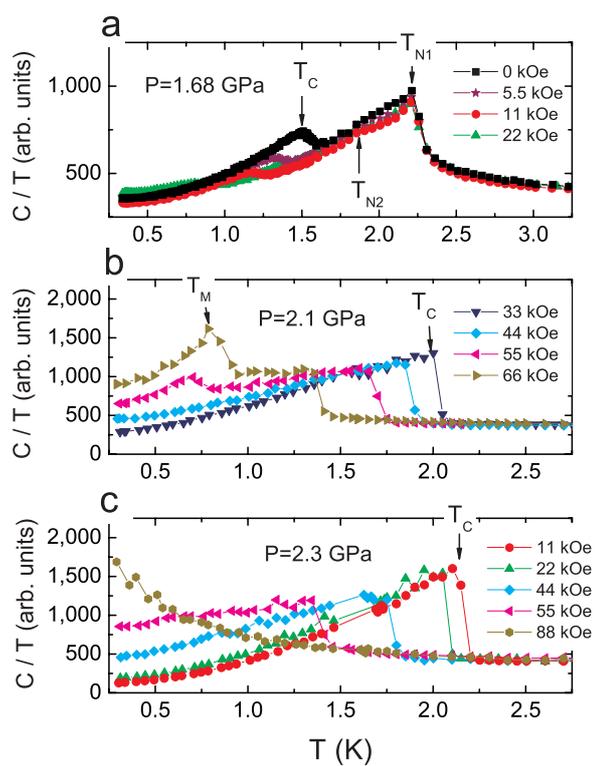

Figure 2

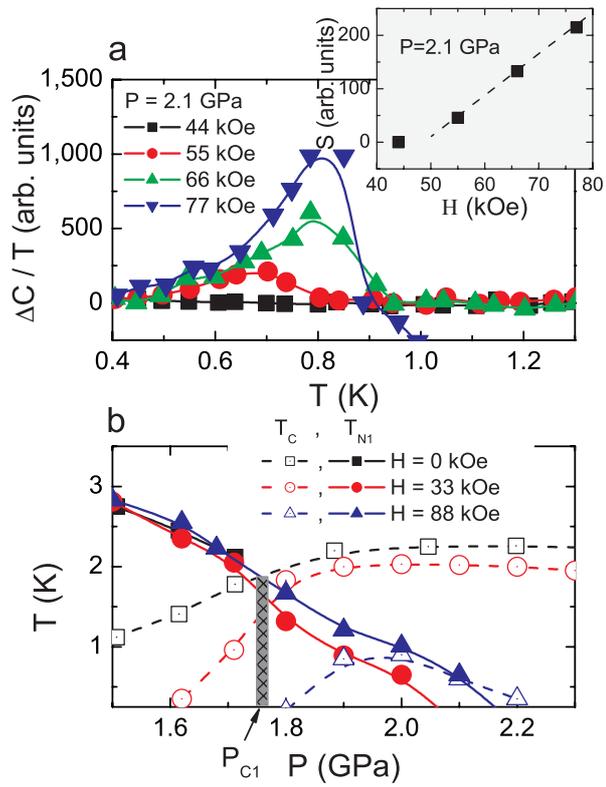

Figure 3

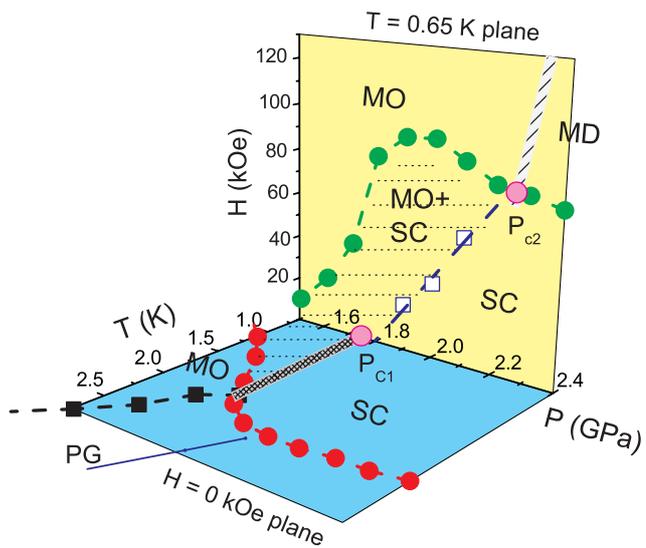

Figure 4

**Supplementary Figure**

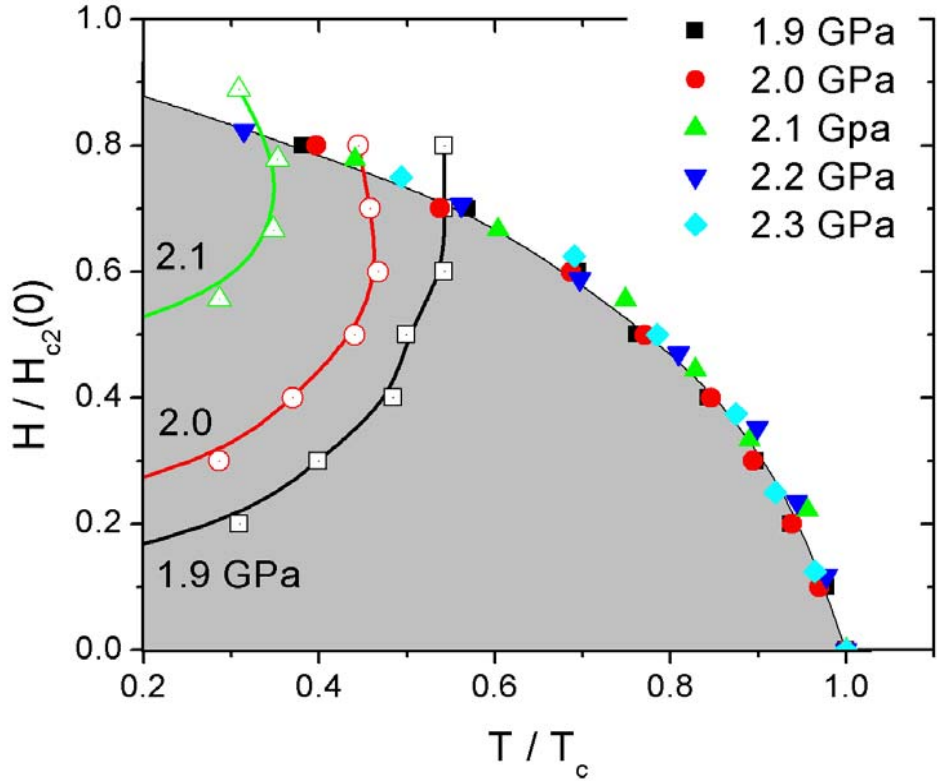

**Figure S1**. *H-T* phase diagram of CeRhIn$_5$ at representative pressures for applied fields perpendicular to c-axis. Solid symbols describe the normalized superconducting (SC) upper critical fields $H_{c2}/H_{c2}(0)$ as a function of normalized temperature $T/T_c$, where $H_{c2}(0)$ is $H_{c2}$ at zero temperature and $T_c$ is SC transition temperature at zero field. Open symbols represent the normalized critical fields $H_c/H_{c2}(0)$ required to induce long-range magnetic order (MO). For $P > P_{c1}$ (=1.75 GPa), a finite field is required to induce long-range magnetic order in the SC state. Solid lines that connect open symbols are guides to eyes and separate a purely superconducting phase from a phase with coexisting MO and SC. For $P$ = 2.3 GPa, only the pure SC phase is only observed for $H \leq 88$ kOe and $T \geq 300$ mK.

## Supplementary Methods: calorimetric measurements under pressure

The measurements of heat capacity are based on an ac calorimetric technique[S1]. Heat is provided by alternating current of frequency $f$, typically 21 Hz, to a heater attached on the back face of a plate-like crystal grown from excess In flux[S2]. The oscillating heat input induces a steady temperature offset $T_{dc}$ from the heat bath with an oscillating temperature $T_{ac}$ superposed. When the measuring time $\tau = 1/f$ is in an optimum range, i.e., $\tau_1 \ll \tau \ll \tau_2$, the oscillating temperature is inversely proportional to heat capacity, $T_{ac} \approx K/C$, where $K$ is a constant that depends on the measuring frequency and the heat input power. The characteristic constants $\tau_1$ and $\tau_2$ are internal sample relaxation and sample-to-bath relaxation times, respectively. The relative value of heat capacity is obtained by converting $T_{ac}$, where $T_{ac}$ is measured using chromel and Au/Fe(0.07) thermocouple wires attached on the front face and is typically 1 mK at $T = 1$ K. The dc offset temperature $T_{dc}$ is kept below 50 mK. The magnetic field dependence of the thermocouple wire is taken into account by using the heat capacity of a nonmagnetic specimen in magnetic field.

For pressure measurements, a hybrid Be-Cu/NiCrAl clamp-type pressure cell with silicon fluid as pressure medium was used to ensure a hydrostatic condition. The value of the pressure at low temperature is determined from the superconducting transition temperature of Sn measured by ac magnetic susceptibility. From the width of the Sn transition, the pressure gradient in the pressure cell is at most ± 0.03 GPa and is independent of applied pressure, reflecting hydrostaticity. Electrical resistivity, measured simultaneously on CeRhIn$_5$ in the pressure measurements, allowed unambiguous identification of magnetic and superconducting transitions.

**Supplementary methods references**